# Managing Demographic Transitions: A Comprehensive Analysis of China's Path to Economic Sustainability

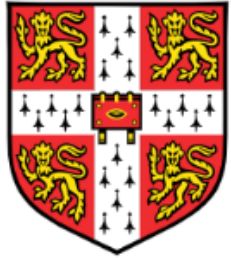

## Hu Yuxin

Faculty of Economics

University of Cambridge

Economics Tripos Dissertation for PartIIB Paper Four:

The Economics of Developing Countries

I'd like to thank Prof Nick Drydakis for his considerate supervision in writing this dissertation. The dissertation is my own work and includes nothing which is the outcome of work done in collaboration expect as specified in the text. It is not substantially the same as any work that has already been submitted before for any degree or other qualification except as specified in the text. It does not exceed the agreed word limit.

**Abstract**

This article presents an analysis of China's economic evolution amidst demographic changes from 1990 to 2050, offering valuable insights for academia and policymakers. It uniquely intertwines various economic theories with empirical data, examining the impact of an aging population, urbanization, and family dynamics on labor, demand, and productivity. The study's novelty lies in its integration of Classical, Neoclassical, and Endogenous Growth theories, alongside models like Barro and Sala-i-Martin, to contextualize China's economic trajectory. It provides a forward-looking perspective, utilizing econometric methods to predict future trends, and suggests practical policy implications. This comprehensive approach sheds light on managing demographic transitions in a global context, making it a significant contribution to the field of demographic economics.

Pembroke College                                                                                             December 2023

# 1. Introduction

In the decades, China has witnessed remarkable economic progress transforming itself from a primarily agricultural society to a global leader, in industry and technology. This transformation has brought along changes in demographics, including an aging population with declining birth rates and substantial migration from rural to areas. These demographic shifts have reaching implications for Chinas labor market, consumer base and overall economic sustainability. Moreover, these trends occur within the context of globalization technological advancements and environmental challenges further complicating the landscape.

Against this backdrop, the research aims to achieve several objectives. Firstly, it seeks to analyze how demographic changes in China impact its growth and development. This involves examining how factors such as an aging population, urbanization and changes in family structure influence labor supply, consumer demand and economic productivity. Secondly, the research aims to assess the effectiveness of policies and strategies implemented by China in response to these demographic shifts. It involves evaluating the role of education, technological innovation and labor market reforms in shaping Chinas trajectory. Additionally, the research poses questions; How have demographic trends impacted Chinas patterns of growth from 1990, to 2050?What is the significance of investing in human capital development to address the challenges posed by an aging workforce? How have advancements, in technology contributed to sustaining growth despite changes? Additionally, what can other countries facing challenges learn from Chinas experiences?

The proposed research holds importance for reasons. Firstly, it enhances our understanding of the relationship between shifts and economic progress a topic that is increasingly relevant in a world with an aging population. Chinas' experience provides insights into managing demographic transitions to support sustained economic growth, which can offer lessons for other nations facing similar circumstances. Secondly, this research sheds light on the effectiveness of policy responses to changes providing a nuanced perspective on how government interventions, economic growth and population dynamics interact. This understanding is crucial for policymakers, economists and international organizations as they navigate challenges stemming from shifts.

To conduct this research, a comprehensive approach is employed that combines analysis with investigation. Theoretical frameworks from schools of thought will be utilized to place Chinas' economic development within a broader economic context. This theoretical foundation will be complemented by an analysis of economic data spanning from 1990 to 2050. The empirical analysis will use econometric methods to examine how demographic factors impact economic indicators. Additionally, we will analyze Chinas' policies to assess their effectiveness in response to changes.

The structure of the thesis will be as follows; In section 2 we will provide a comprehensive review on the current economic theories in terms of economic development, particularly focusing on the relation to the Chinese context. We will discuss the Classical, Neoclassical and Endogenous Growth theories and their implications for understanding Chinas development. Section 3 will provide a review of the relationship between Chinas' population and economic growth from 1990 to 2020. Section 4 will analyze economic data from China including trends in population dynamics, labor market changes and patterns of economic growth. In section 5, based on the previous evaluations, we will formulate our policy implications. Finally, in section 6, we will conclude the paper by summarizing findings discussing policy implications and suggestions for further research.

# 2. Theoretical considerations

In Section 2 we delve into an in-depth analysis of economic theories that have examined the connection between demographic trends and economic development. We start by exploring the Classical School, which was shaped by thinkers like Malthus, Smith and Ricardo. Their ideas shed light on how population growth can pose limitations on economic resources. Moving forward we then shift our focus to the Neoclassical School and its Solow Growth Model. These models provides insights into the roles played by capital, labor and technological progress in economic growth. Additionally, we explore the Endogenous Growth Theory, which emphasizes the importance of innovation and knowledge for economic progress through models developed by Kremer and Galor as well as Weil. By internalizing advancements within this theory, it offers a more comprehensive understanding of Chinas complex economic landscape. Lastly, we also consider Barro and Sala i Martins (1995) model, for our further analysis in section 4.

2.1 The Classical School

The classical school of thought popularized by figures such as Adam Smith, David Ricardo and Thomas Malthus emphasized the strong connection between population growth and economic development. Malthus (1798) argued that an important aspect of this perspective is the belief that if population growth remains unchecked it will eventually outpace agricultural production capacity leading to famine and increased mortality rates. Moreover Ricardo (1817) elaborated on this idea by introducing the 'Law of Diminishing Returns' which suggests that as population increases the additional output resulting from labor and land would gradually decrease, thus limiting prosperity. While Adam Smith (1776) held a pessimistic view he acknowledged that expanding populations could strain resources, he also recognized the potential, for market mechanisms and capital accumulation to help alleviate these challenges.

Firstly, Thomas Malthus presents a grim scenario driven by the law of diminishing returns vis-à-vis a growing population. This disparity, as illustrated through the Malthusian population function, aligns population growth with the vicissitudes of real per capita income, dependent on birth and death rates. The model concisely captures the essence of resource scarcity, with a production function $Y = f(L, N)$, where output Y is determined by the interplay between labor L and a fixed amount of arable land N. As labor grows, the plague of diminishing returns kicks in due to the diminishing per capita land, causing a deceleration in economic growth (Van den Berg, 2017). Classical model usually depicts that any technological improvements that augments the average product of labor will, in the long term, be neutralized by population growth and the consequent diminishing returns (Van den Berg, 2017). As demonstrated by figure 1 , if the number of workers increases, total output grows but at a diminishing rate relative to population growth, leading to a decrease in per capita output, while a decrease in workers results in higher per capita output, with population level PA acting as a stable equilibrium point (Drydakis, 2022).

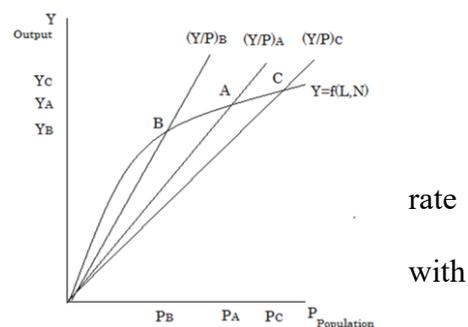

Firgure 1

The Malthusian model's premise of a direct correlation between population growth and resource limitations has manifested in various regions and historical periods. For example, during the pre-industrial periods, Europe exhibited Malthusian dynamics where periods of population growth were often followed by survival crises caused by failings of agricultural output to keep the pace (Clark, 2007). Similarly, China during its Qing dynasty also experienced cycles of population expansion and contraction following the fluctuations of the agricultural productivity (Lee, 1987). Meanwhile, some sub-Saharan African countries facing rapid population growth faced scarce resources echo exactly described by the Malthusian scenarios (Dasgupta, 1993). These instances underscore the model's relevance in specific socio-economic contexts, but notably, it only signifies where technological and industrial advancements have yet to significantly alleviate resource constraints.

The Malthusian model has received criticism, for its assumptions given the advancements in technology observed from the 18th to the 21st century. One of its key drawbacks is the failure to adequately consider how technological progress can alleviate resource constraints and support both population and economic growth (Van den Berg, 2017). This becomes more evident when we look at the agricultural and industrial innovations that have greatly expanded economies capacity to sustain growth. At the time, the emergence of Growth theory has transformed economic thinking by emphasizing the importance of knowledge, technological change and human capital in fostering economic growth. This challenges the claims about stagnation caused by resource limitations from the classical school thoughts.

## 2.2 The Neoclassical School

The Neoclassical school of thought focuses on how the combination of capital accumulation growth in the labor force and technological progress interact to shape the long-term growth path of an economy (Van den Berg, 2017). Unlike the model presented in the classic school, the neoclassical framework suggests that economies can sustain both population and economic growth by increasing the investments in capital and advancing technology. However, it also states that in the long run, per capita economic growth is primarily driven by external technological progress rather than population growth. Additionally, the Neoclassical school recognizes that rapid population growth could potentially decrease per capita income levels if not balanced with investments in capital and technology (Barro & Sala i Martin, 2004). Overall, from a perspective we gain a comprehensive analysis and a more optimistic understanding of how population dynamics influence economic growth by considering factors such, as capital accumulation, technological innovation and human capital development.

The Solow Growth Model as a subset of the Neoclassical School posits that technological progress is an essential driver of continuous economic growth, thereby offsetting the effects of diminishing returns on capital (Solow, 1956). Particularly, the model delineates how technological growth, denoted as $\frac{\Delta A}{A} = g$, interacts with capital K and labor L inputs to influence real GDP Y growth. The pivotal equation of the model is expressed as $\frac{\Delta Y}{Y} = g + a\left(\frac{\Delta K}{K}\right) + (1-a)\left(\frac{\Delta L}{L}\right)$ where a is the share of capital in output. In addition, technological progress galvanizes growth in output per worker at a rate exceeding the rate of technological progress itself, due to the concomitant capital accumulation spurred by enhanced output per worker. This dynamic is succinctly captured in the equation $\frac{\Delta y}{y} = g + a\left(\frac{\Delta k}{k}\right)$ where y and k denote the real GDP and capital stock per worker, respectively (Solow, 1956). In the steady-state realm delineated by Solow, the growth rates of real GDP per worker and capital per worker are harmonized, encapsulated by the equation $\frac{\Delta y}{y} = \frac{g}{1-a}$ (p.49, Drydakis, 2022) where y and k represent the real GDP and capital stock per worker, respectively. This steady-state growth underscores a scenario where the growth impetus from technological progress balances the diminishing returns of capital, fostering a long-term growth equilibrium.

The Solow Model is often regarded as a reflection of how economic growth works in regions and time periods. For example, it aligns with the World War II economic boom, in Western countries like the United States and parts of Europe, where capital deepening and technological innovation played crucial roles (Solow, 1956). Moreover, the model has been utilized to understand the economic "miracles" of East Asian economies like South Korea and Taiwan during the latter half of the 20th century, where substantial investments in capital and a growing labor force contributed to rapid economic expansion (Young, 1995). Furthermore, the model can explain the paths taken by India, where capital investments, labor force growth and technological progress have been drivers of their growth (Bosworth & Collins 2008).

The Solow Growth Model also exhibits some limitations concerning the treatment of population growth. First, it assumes a constant rate of population growth without considerring any feedback mechanisms between economic growth and demographic changes (Galor & Weil 2000). Economic prosperity may affect population growth rates through alterations in fertility and mortality rates. This is a complexity not captured in the Solow framework. Secondly, the model oversimplifies the workforce by treating all workers as homogenous, disregarding the varying levels of skills, education and productivity, among individuals (Lucas, 1988). This is a significant limitation given that human capital development has been identified as a crucial factor for economic growth. Additionally, the Solow model assumes that technological progress is exogenous and unaffected by population growth or density, whereas in reality, larger or denser populations might foster innovation and technological advancements through various channels like knowledge spillovers or larger markets for goods and ideas (Romer, 1990).

## 2.3 Endogenous theory: Transcending the Limitations of Exogenous Technological Progress

The endogenous growth models which explore the mechanics of change are frequently considered to be an extension of the Solow model. However, it is more accurate to view these models as complementary, than replacing the Solow model, as they elucidate the underlying dynamics of technological change, which the Solow model treats as exogenous. Unlike the Solow model, which sees technological progress as an exogenous factor, the Endogenous Growth Theory suggests that economic factors and policies play a significant role in determining the pace of technological innovation and thus influencing growth trajectories (Zhao, 2018).

Endogenous Growth Theory emphasizes that innovation and technological progress are primarily driven by profit seeking entities. According to this theory as human knowledge accumulates, the cost of innovation decreases. Unlike the Solow model, which assumes diminishing returns to capital, endogenous growth models propose that, under certain conditions, there can be non-diminishing or even increasing returns to capital, especially when combined with capital and knowledge accumulation. This notion is substantiated by the Grossman and Helpman (1990) model, where the growth rate of new products, denoted by g, is expressed as $g = \frac{L(1-\alpha)}{c - \alpha r}$, illustrating that a larger stock of individuals engaged in research and development L propels a higher rate of growth. $L(1 - \alpha)$ suggests that the amount of labor L dedicated to research and development, adjusted by the factor $1 - \alpha$, has a direct impact on the growth rate of new products g. It can be inferred that as the amount of labor in R&D increases, the growth rate of new products also increases. Furthermore, $c - \alpha r$ represents the adjusted cost of invention. Here "c" represents a denoting the initial cost of invention while $\alpha r$ symbolizes a representation of discounted costs over time or, under varying economic conditions. The division between these two parts indicates that the rate of growth for products denoted as g is determined by the ratio of the adjusted labor input to the adjusted cost of invention.

2.3.1 Kremer model

Kremers study examines the relationship between population growth and technological progress from a perspective that diverges from traditional Malthusian theories. Of focusing solely on a dual-faceted role of population growth, Kremer highlights not only the conventional congestion effects but also the creative effects that potentially fuel technological innovation. In his model, he introduces the equation $\Delta A = qPA$, which morphs into $\frac{\Delta A}{A} = g_A = qP$, asserting that technological growth rate is directly proportional to population size, given a constant research productivity per individual. This aligns with the proposition that a burgeoning population, through a lot of minds, cultivates fertile soil for innovative ideas. Kremer extends this narrative into a production function $Y = AP^a N^{\{(1-a)\}}$ and the per capita income expression $\frac{Y}{P} = AP^{a-1}N^{1-a}$, which reminds of Malthusian thought, where an increase in population, with other conditions remaining unchanged, leads to a decline in per capita income. However, Kremer's departure from Malthus lies in the assumption that population growth, spurred by per capita income surpassing a subsistence level $y_s$, also propels technological progress. The long-run steady state of population $P^* = ZA^{\frac{1}{1-a}}$ as postulated by Kremer, is tied to the level of technology A, implying at a generic relationship between technological advancement and populational size. This notion gains further trcation with the expression $g_{\{P^*\}} = \frac{q}{1-a}P$, linking population growth rate to research producctivity and the relative signficance of natural resources in the production function.

Kremer's exploration into this two-way connection brings together the contradictory perspectives on congestion and creativity in relation to population growth. This provides a perspective for examining how technological progress and population growth interact. Nonetheless, Kremer's model encounters limitations when juxtaposed with empirical observations of per capita output growth over the past two centuries and the recent slowdown in population growth. Such limitations highlight areas that can be further refined and explored in the model.

2.3.2 Galor and Weil model

The model created by Galor and Weil offers insights into the relationship between growth and population dynamics across different historical periods. They classify the development into three stages; the Malthusian, Post Malthusian and Modern Growth phases, where they explore the intricate connections between population growth, technological advancements and economic progress. A crucial element of their framework is incorporating the internalization of progress and population growth rates as elements, within the broader economic context.

According to Galor and Weil, the concept of progress denoted as $g_A$ is linked to the population's human capital. They express it as a function of the capital augmented population represented by $g_A = f(HP) = f(Q)$ where P signifies the size of the population H represents the index for human capital per person and Q is calculated as HP. This formulation emphasizes the posilitive relationship between a populations human capital environment and the rate of technological advancement, which forms a fundamental aspect of their model.

Galor and Weil further elaborate on this idea by suggesting a self-reinforcing cycle in which technological progress enhances returns on investments in capital. These increased returns serve as motivation for investments in human capital per child supported by rising incomes. As a result, there is an expansion, in the capital augmented population. This expansion then fuels advancements in technology creating a beneficial cycle of growth.

The Modern Growth phase, as delineated by Galor and Weil, emerges as parents, buoyed by rising per capita incomes and elevated returns on human capital, opt for a strategy of "quality over quantity" in child-rearing. They invest more in the human capital of each child while having fewer children, a dynamic that not only propels the human-capital-augmented population but also accelerates technological progress. With population growth rate moderating and technological progress surging, the phase manifests in an accelerated growth of per capita output.

In the modern contexts, we can observe the Kremers model in action in developing nations like China and India which have large populations. The size of these regions populations has been linked to the rise of innovations and economic advancements from the late 1900s to the early 2000s (Zhang, 2015). For instance, Chinas significant population has been identified as a driving force behind its progress and subsequent economic success on a global scale. On the hand the Galor and Weil model is particularly relevant for countries going through demographic changes. As an example, this model helps us understand how several Asian economies such as South Korea and Taiwan experienced shifts in their population demographics during the 20th century. These nations witnessed declining fertility rates alongside an increase, in their educated workforce. With technological advancements, these factors propelled them into sustained periods of economic growth (Galor, 2011).

The Kremer and Galor and Weil models, while insightful, bear certain limitations. Kremer's model tends to oversimplify the multifaceted nature of inovation processes. It overlooks factors like institutional frameworks, policy environments, and global collaboration which significantly influence technological advancements, irespective of population size (Jones, 2005). Moreover, the model assumes a linear progression of innovation with population growth, which might not hold in scenarios of resource constraints or adverse environmental impacts associated with overpopulation. On the other hand, the Galor and Weil model, though robust in explaining the transsition from Malthusian stagnation to modern growth regimes, might not fully account for disparities in growth trajectories among different regions (Strulik & Weisdorf, 2008). The model's assumption of a universal process of demographic transition and economic growth can oversimplify the diverse socio-economic and cultural context across the regions. Aditionally, the model's emphasis on human capital accumulation as a driver of growth, while valid, may not capture other crucial factors like political stability, economic policies, and global economic dynamics that significantly impact a nation's growth trajectory (Prettner, 2014).

2.4 Barro and Sala-i-Martin (1995) Model

In the study conducted by Barro and Sala i Martin in 1995, they put forth an intricate economic growth model that primarily focuses on the convergence hypothesis. This model, rooted in growth theory suggests that less prosperous economies tend to experience faster growth rates compared to wealthier ones ultimately leading to a narrowing of income disparities over time. Barro and Sala i Martins (1995) research delves into this convergence phenomenon through both country to country and region, to region comparisons utilizing a dataset and advanced econometric techniques.

Their model focuses on the concept of " convergence," which suggests that economies tend to reach stable levels of per capita income when they share similar characteristics like savings rates, population growth and investments in human capital. This perspective differs significantly from the idea of convergence and offers a more realistic framework for understanding patterns of economic growth. To analyze this, they use regression techniques to examine the relationship between the growth rate of per capita GDP and initial levels of per capita GDP. They consider factors such as investments in human and physical capital, government consumption, fertility rates and market distortions. The results show a correlation between initial income levels and subsequent growth supporting the theory of conditional convergence. Additionally, Barro and Sala i Martin (1995) expand their analysis beyond borders to explore regional convergence within countries. This aspect emphasizes the role played by factors, in economic growth and highlights that convergence is not exclusively a global phenomenon but also occurs within national boundaries. Furthermore, Barro and Sala i Martin (1995) extend this analysis to investigate convergence within countries. This approach entails utilizing regression models but with a specific emphasis, on analyzing different areas within a single country. The objective is to investigate whether affluent regions demonstrate a tendency to narrow the gap and catch up with more prosperous regions over a period of time.

A central part of their quantitative analysis is the estimation of a convergence equation, typically in the form of:

$$\text{Growth Rate}_{i,t} = \alpha + \beta \times \log(\text{GDP Per capita}_{i,t-1}) + \gamma' X_{i,t-1} + \epsilon_{i,t}$$

Here, Growth Rate$_{i,t}$ represents the average annual growth rate of per capita GDP for country or region I over a period t. The term log (GDP Per capita$_{i,t-1}$ is crucial, as it captures the initial level of income, with the expectation (based on the convergence hypothesis) that $\beta$ would be negative, indicating that countries with lower initial GDP per capita tend to grow faster.

The vector $X_{i,t-1}$ includes other explanatory variables like investment rates, population growth, government expenditure, and measures of human capital, among others. These variables are crucial for the concept of "conditional convergence," as they account for differences in steady-state income levels across countries or regions. Their findings typically show a significant negative coefficient for log (GDP Per capita$_{i,t-1}$), supporting the conditional convergence hypothesis. This also means that, after controlling for various factors, poorer economies grow faster than richer ones, albeit converging to different steady-state levels.

However, there are some limitations to the model to be considered. One significant criticism is that it may oversimplify economic realities. While it considers the levels of development through conditional convergence, it might not fully capture the complexities of each countrys' unique historical, institutional and cultural contexts. This gap can limit its applicability in situations especially in economies where non-economic factors significantly impact growth patterns (Durlauf & Quah 1999). Moreover, we can question the assumption of technology across regions, which is a fundamental premise of the model. In reality, technology diffusion tends to be uneven and influenced by factors, beyond mere economic indicators (Acemoglu & Robinson 2012).

3. A Chronological Review of The Relationship Between China's Population and Economic Growth

According to Liu and Lu (2019), they divide the aging trend in China into four stages: the transition phase (1990-2005) and the rapid aging phase (2006-2020). Different theories are applicable in the different phases.

3.1 Transition Stage

The window of opportunity for China began in 1990 when the total dependency ratio dropped to 49.8%. A lower dependency ratio indicates a higher proportion of working-age individuals, leading to a larger labor supply in the job market. It also means there are fewer people to support, allowing for more savings to be accumulated for productive economic investments. Additionally, smaller family sizes permit greater investment in the education and health of children. This can boost worker productivity and possess valuable human capital.

This period of rapid economic growth coincided with the opening of China's demographic window of opportunity. This implies that the economic growth in China may be connected to changes in the population structure. What makes it more intriguing is that, since the 1980s, China has introduced reforms aimed at transitioning its economy into a market system. These reforms were implemented to improve the allocation of economic resources with a particular focus on eliminating barriers for private sector development (Chen and Feng 2000). The market-oriented reforms significantly increased flexibility in labor and capital markets allowing for absorption of the larger workforce entering the job market during the demographic window. Additionally, these reforms facilitated investments by converting accumulated savings into economic growth. This indicates that economic reforms those focused on market orientation might have created an enabling policy environment for China to benefit from its demographic dividend. As shown in figure 2, China's economic freedom index exhibited a consistent increasing trend from 1990 to 2005 ( Gwartney et al. , 2015).

At this stage, even though China has entered the transition stage of aging, it is still in the early phase of population aging. A study conducted by Liang et al. (2023) reveals that, due to the growth of both the working age population and the elderly population with the former growing faster than the latter, there exists a positive correlation between population aging and per capita GDP. This perspective challenges the held belief among researchers, such as Maestas et al. (2014), that aging has detrimental effects on economic development. Liangs' research along with that of Yang et al. argues against using linear models to study how population aging affects growth as this approach overlooks crucial factors such as the relative proportions of different age groups within the population and their contributions to the economy. During the stage of population aging, although there is an increase in elderly individuals, there is also a rise in young individuals (aged 15 64). Consequently, at this stage any negative impact on growth caused by an aging population will be counterbalanced by positive contributions, from younger generations. Therefore, in the early phase of population aging, aging will promote economic growth.

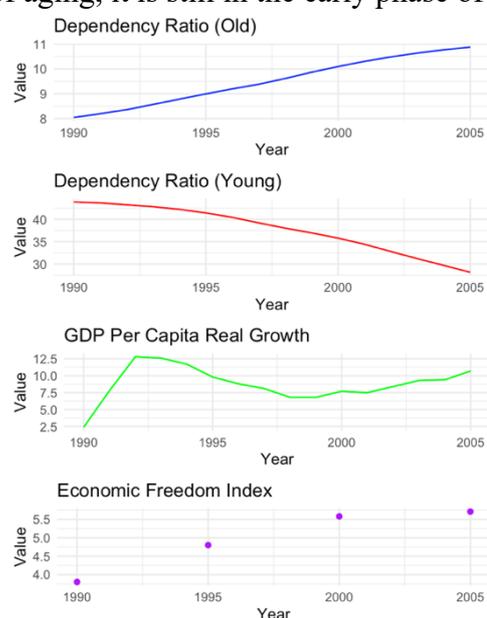

Figure 2[1]

Using the Solow growth model as a framework, we can delve deeper into how Chinas economic growth relates to its changes between 1990 and 2005. The Solow model highlights the significance of accumulating capital growing labor force and technological progress in driving growth. During this period Chinas' demographic advantage played a role in boosting its labor force. As demonstrated in figure 2, the country witnessed a decline in its dependency ratio resulting in a larger proportion of working age population. According to the Solow model, this contributed positively to output. Moreover, figure 3 showed the increased savings rate, another key variable in the Solow model, can be attributed to the reduced family sizes and the need to support fewer dependents. These accumulated savings provided China with the capital for investments in infrastructure and technology further fueling economic growth. Additionally, with increased investments in education (figure 4), the quality of the labor force improved, which can be interpreted in the Solow framework as a form of human capital enhancement. This, in turn, could lead to higher steady-state levels of output per worker.

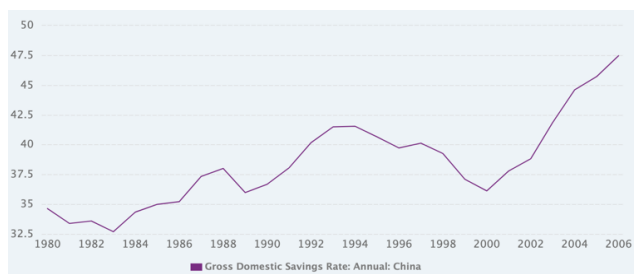

Figure 3[2]

---

[1] Source: OECD Labour Force Statistics 2022, https://doi.org/10.1787/dc0c92f0-en

[2] Source: World Bank Indicators. 2023

The market-oriented reforms implemented by China can be seen as a mechanism to optimize the allocation of these crucial resources — capital and labor. In the context of the Solow model, these reforms would help in converging faster to the steady-state, maximizing the benefits derived from the demographic dividend. The model underscores the importance of maintaining a balance between labor, capital, and technology for sustainable growth. With China's demographic changes acting as a catalyst, the market-oriented reforms ensured that the country efficiently utilized its available resources, aligning perfectly with the predictions of the Solow growth model.

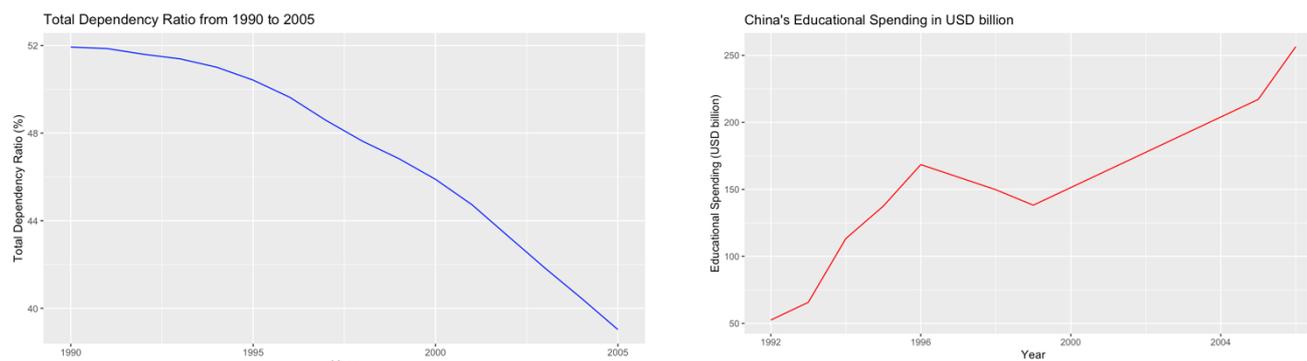

Figure 4[3]

3.2 The Rapid Aging Phase

At this stage, China formally entered an aging society. During the accelerated phase (2001-2018), the first wave of baby boomers born shortly after the establishment of the People's Republic of China in 1949 became senior citizens. In this phase, the annual growth rate of the population aged 65 and above reached 3.28%, significantly exceeding the 0.66% annual growth rate of the total population (Lu & Liu, 2019). By 2014, the populations aged 60 and 65 accounted for 15.53% and 10.06% respectively, and the dependency ratio calculated based on the working-age population (ages 15 to 59) began to decline in the same year. Additionally, the growth rate of China's working-age population decreased, and in 2015 the absolute number of working-age adults began to decline, leading to a structural labor shortage (Fang & Wang, 2006). Meanwhile, China also faced pressures brought about by economic development. To tackle issues arising from an aging population and labor shortages caused by these shifts in demographics and economic development pressures; China decided to end its one child policy after three decades or more in effect. In its place came a selective two child policy implemented in 2013 that allowed couples, from one child families to have a child if either spouse met that criterion. Three years later, in 2016, another policy change allowed all couples to have a second child (i.e., the comprehensive two-child policy). Although some studies believe that China's two-child policy can effectively promote national economic growth, others argue that the policy can only play a limited role in this regard (Li, 2016).

From 2006 to 2020, China underwent profound changes in its economic landscape, characterized by an accelerated pace of marketization and a series of significant reforms in its economic institutions. These transformations were driven by the shift from a largely planned economy to a more market-oriented one, fostering greater economic efficiency, dynamism, and competitiveness on the global stage (Zhang & Li, 2018). At the time, China's financial sector underwent notable liberalization, with regulatory reforms aiming to foster innovation, improve financial stability, and better serve the real economy (Chen & Yao, 2016).

---

[3] Source: World Bank Indicators. 2023

Moreover, to further maintain sustainable growth, China introduced reforms in areas including state-owned enterprises, land, labor, and fiscal systems (Lin & Appelbaum, 2016). The progress China has taken in this period have played a pivotal role in its economic rise and integration into the global economic system.

China also witnessed a transformative phase in its educational sector during the period, which has played a pivotal role in its human capital development. The country made substantial investments in primary, secondary, and tertiary education, resulting in increased enrollment rates and improved educational outcomes (Liu & Wang, 2015). This growth in education was bolstered by policy initiatives such as the "National Outline for Medium and Long-term Education Reform and Development (2010-2020)," which aimed to cultivate a talent-driven economy and bridge the urban-rural education divide (Wang & Huang, 2012). Moreover, the emphasis on vocational education and training (VET) has enabled China to better align its labor force with the demands of its rapidly industrializing economy (Zhang & Zhao, 2017). The expansion of higher education institutions, coupled with international collaborations, has also fostered a culture of research and innovation, leading to an uptick in global academic publications and patents (Li & Zheng, 2019). Collectively, these educational advancements have bolstered China's human capital, positioning it as a key player in the global knowledge economy.

China, in the meanwhile, experienced a meteoric rise in technological advancements, solidifying its position as a global powerhouse in the tech arena. This period was marked by substantial state-driven investments in research and development (R&D), leading to breakthroughs in various fields ranging from artificial intelligence to biotechnology. The "Made in China 2025" initiative, launched in 2015, epitomized the nation's ambition to transition from a manufacturing-centric economy to one rooted in innovation and high-tech industries (Wu & Chen, 2017). Furthermore, the proliferation of tech giants like Alibaba, Tencent, and Huawei not only transformed domestic digital landscapes but also expanded China's technological influence globally (Yang & Clarke, 2019). Collaborations between academia and industry, bolstered by policy incentives, have fostered a conducive environment for startups, leading to the emergence of a vibrant tech ecosystem in cities such as Shenzhen and Hangzhou (Li, 2018). This rapid technological ascent has had profound implications, reshaping both China's domestic economic fabric and its stature in the global technological hierarchy.

Examining Chinas trajectory from 2006, to 2020 using the Galor and Weil model provides insights into the nation's connection between demographics and economy. According to Galor and Weil during the Modern Growth phase countries emphasize quality over quantity in raising children leading to investments in capital despite a slowdown in population growth. Chinas experience during this period reflects these principles. As mentioned earlier China witnessed a rise in its aging population while the working age group started declining after 2015. In response to this shift and with an educated populace China focused heavily on developing human capital by making significant advancements in primary, secondary and tertiary education. This emphasis on education deviates from Galor and Weils model of prioritizing capital driven population growth but also highlights Chinas shift. By augmenting investments in capital amidst an aging population China aimed to leverage the "quality" of its people to drive economic progress. The resulting surge in advancements and economic reforms during this period can be partially attributed to this focus, on human capital aligning with Galor and Weils theories of Modern Growth phase. Hence, while the challenges of an aging demographic are undeniable, China's strategic human capital investments during this period, as reflected in the Galor and Weil framework, underscore its efforts to transform demographic challenges into economic opportunities.

4. Empirical Analysis

In investigating the long-term economic implications of demographic shifts within China's workforce, we draw upon the theoretical underpinnings of the demographic-economic relationship as suggested by the seminal work of Barro and Sala-i-Martin (1995) and extend the model to incorporate unique demographic transitions experienced by China since 1990. Our econometric model is constructed to evaluate the first and second order demographic effects posited to influence economic output per capita, specifically focusing on the age cohorts of 15–34 and 35–54, which represent young and prime working-age populations, respectively.

The data presented span a period from 1990 to 2050, covering a range of economic indicators across various provinces in China. The first period (1990-2020) relies on global and national data sources, notably the United Nations' World Population Prospects: The 2022 Revision, offering a comprehensive view of population dynamics. Migrant stock trends are sourced from the United Nations Department of Economic and Social Affairs (2023), ensuring a detailed understanding of migration patterns. The UNESCO Institute for Statistics (UIS) provides educational statistics, enhancing the depth of socioeconomic analysis. In the context of China, provincial level panel data from 31 major provinces form the basis for evaluating economic and demographic shifts. These datasets include economic indicators and population figures segmented by age and employment status, offering a granular view of China's evolving economic landscape.

Looking ahead to 2020-2050, the study leans on the datasets from the research of Chen et al. (2020), focusing on provincial and gridded population projections for China under shared socioeconomic pathways. Chen et al. (2020) provided a year-by-year prediction on China's demographic structure for each of the thirty-one provinces. Their data is based on these assumptions: The rivalry between nations will lead to the formulation of national policies centered on security. In China, a fully liberalized fertility policy will lead to significant population growth, helping to better preserve human resources. However, minimal investments in education and healthcare will lead to increased mortality rates and lower levels of education. Despite the worsening inequality, stagnant economic growth across all provinces will reduce the movement of people between these regions. This forward-looking data allows for the analysis of future demographic trends and their implications for economic and social planning. To address future economic indicators for China, a potential source could be the Asian Development Bank's annual Asian Development Outlook, which provides comprehensive analyses of economic and development issues in Asia and the Pacific, including projections of key economic indicators for the region's economies. This would offer a well-rounded perspective on China's future economic trajectory, grounded in rigorous research and analysis.

A comprehensive list of data sources this paper has utilized is in section 7/

We specify our econometric model as follows:

$\Delta \log GDP_{it} = \alpha + \beta_1 \log GDP_{it-1} + \beta_2 YoungAgeRatio_{it} + \alpha_{it} + \varepsilon_{it}$  I

$\Delta \log GDP_{it} = \alpha + \beta_1 \log GDP_{it-1} + \beta_2 PrimeAgeRatio_{it} + \alpha_{it} + \varepsilon_{it}$  II

$\Delta \log GDP_{it} = \alpha + \beta_1 \log GDP_{it-1} + \beta_2 YoungAgeRatio_{it} + \beta_3 PrimeAgeRatio_{it} + \alpha_{it} + \varepsilon_{it}$  III

$\Delta \log GDP_{it} = \alpha + \beta_1 \log GDP_{it-1} + \beta_2 YoungAgeRatio_{it} + \beta_3 PrimeAgeRatio_{it} + \beta_4 YoungAgeRatio_{it} * \mathbf{PeriodDummies} + \beta_5 WorkingAgeRatio_{it} * \mathbf{PeriodDummies} + \alpha_{it} + \varepsilon_{it}$  IV

$\Delta \log GDP_t = \alpha + \beta_1 \Delta \log GDP_{t-1} + \beta_2 YoungAgeRatio_{it} + \beta_3 PrimeAgeRatio_{it} + \beta_4 YoungAgeRatio_{it} * Educational_{Attainment1} + \beta_5 PrimeAgeRatio_{it} * Educational_{Attainment2} + \alpha_{it} + \varepsilon_t$  V

$\Delta LogRD_t = \alpha + \beta_1 WorkingPopulationRatio_t + \beta_2 TertiaryEducationEnrollment_t + \alpha_{it} + \epsilon$  VI

$\Delta \log GDP_t = \alpha + \beta \Delta LogRD + \alpha_{it}$  **VII**

Here, $\Delta \log GDP_{it}$ is the growth rate of the logarithmic real GDP per capita for China in province i at time t, capturing the dependent variable of economic growth. The term $\log GDP_{t-1}$ is included to account for conditional convergence effects. The variables $YoungAgeRatio_{it}$ and $PrimeAgeRatio_{it}$ represent the independent variables of interest: the proportion of the population within the working age (15–64) and prime age (35–54) brackets, respectively. $X_t$ is a vector of control variables, such as capital stock, human capital, and technological progress, which may affect economic growth. PeriodDummies has three values: 0=time period from 1990 to 1999, 1=time period from 2000 to 2009, 2=time period from 2010 to 2022. $\alpha\_it$ is provincial fixed effects.

When interacting with educational attainment level of data, due to constraint of data that average years of school by age group could not be obtained, we interact tertiary school enrollment (the ratio of total enrollment to the population of the age group that officially corresponds to the level of education shown) with ratio of young age workers. The tertiary school enrollment will be lagged by three periods to account for the lagged effect of school enrollment on economic growth to obtain Educationa_Attainment1. For the prime workers, we will still use tertiary school enrollment but lagged by 13 periods, as a worker aged 35 will have completed 13 years earlier of tertiary education with Educationa_Attainment2. We will interact secondary school enrollment as well in a similar way. Please note that due to the Cultural Revolution, there was not GaoKao (Chinese national tertiary school entrance school exam) from 1982 to 1983. We will put zeros into these two years.

Model (vi) and (vii) aims to examine how demographic factors, specifically the proportion of the educated population and the working-age population, influence technological advancements. A positive coefficient for the proportion of the educated population would suggest that a higher level of education (especially tertiary education) is associated with greater technological advancements. Due to data constraint, the analysis is reframed within 1996 to 2020. TertiaryEducationEnrollment_t is gross enrollment ratio is the ratio of total enrollment, regardless of age, to the population of the age group that officially corresponds to the level of education shown. $\Delta LogRD\_t$ stands for growth rate of logged research and development expenditure.

The model has considered Barro & Sala-i-Martin (1995) model and the concept of economic inertia, as described in the field of macroeconomics, posits that the economic growth of a country in a given year is often influenced by its performance in the preceding periods. phenomenon is rooted in the persistence of economic conditions, such as capital accumulation, technological advancements, and labor force trends, which do not immediately adjust to new equilibriums (Barro & Sala-i-Martin, 2004). By including logGDP_(t-1), the model captures this temporal continuity, providing a more comprehensive understanding of the growth dynamics.

Moreover, the inclusion of the lagged GDP term infact aligns with the theory of autoregressive processes in time-series analysis, where past values of a variable are used to predict its future values. This approach also acknowledges the path-dependent nature of economic development, pointed out by Hamilton (1994). The persistence of economic growth, as captured by the lagged GDP term, in the meanwhile, also reflects the underlying structural characteristics of the economy, including institutions, policy continuities, and long-term investments, which are crucial in understanding economic paths(Acemoglu & Robinson, 2012).

Province fixed effects: we include provincial dummies in the regressions.

## 4.1.1 Empirical Examinations of China's Demographic Structure's Impact on Economic Growth from 1990 To 2020

The results of analysis is summarized in the following table:

| | | GDP Growth Rate (Logged) | | | | | | R&D Expenditure (Logged) |
|---|---|---|---|---|---|---|---|---|
| Model | (i) | (ii) | (iii) | (iv) | (v) | (vii) | | (vi) |
| log(GDP Per Capita_(t-1)) | -0.4349***(0.002) | -0.3672**(0.08) | -0.5605***(0.0001) | -0.2762*(0.064) | -0.4675*(0.08) | -0.3765*8(0.049) | | |
| YoungAgeRatio_it | 0.1365***(0.003) | | 0.4589**(0.033) | 0.78889**(0.043) | 0.52688**(0.055) | | | |
| PrimeAgeRatio_it | | 0.1533**(0.028) | 0.7561**(0.022) | 0.9765**(0.039) | 1.13788**(0.064) | | | |
| PeriodDummies1*YoungAgeRatio | | | | 0.5582(0.145) | | | | |
| PeriodDummies 2*YoungAgeRatio | | | | 1.48441(0.197) | | | | |
| PeriodDummies 3*YoungAgeRatio | | | | -0.37074(0.156) | | | | |
| Educational_Attainment1_t* YoungAgeRatio_it | | | | | 0.09553**(0.03542) | | | |
| Educational_Attainment2_t*PrimeAgeRatio_it | | | | | -0.07949*(0.067) | | | |
| R&D Expenditure (Logged) | | | | | | 0.667359***(0.001) | | 10.8801***(0.009) |
| WorkingAgePopulationRatio)_it | | | | | | | | 26.5122***(0.00619) |
| Province fixed effects | Yes | Yes | Yes | Yes | Yes | Yes | | Yes |
| Adjusted R^2 | 0.367 | 0.423 | 0.584 | 0.567 | 0.643 | 0.5985 | | 0.8934 |

Notes: All results are OLS estimates. ***, **, and * indicate statistical significance at 1%, 5% and 10%, respectively. The p-values are included in the brackets.

Across models I to VII, the coefficient for lag_logGDP is all negative and statistically significant, which is consistent with the Barro & Sala-i-Martin (1995) model's prediction illustrated in section 2.4.

In model I, the YoungWorkersRatio coefficient is $\beta_1 = 0.1365$, significant at the p=0.003 level, indicating a positive relationship with economic growth. This suggests that a higher proportion of young workers in the workforce is associated with higher economic growth. This aligns with the theoretical perspective that young workers often bring innovation, adaptability, and a dynamic skill set, fostering an environment conducive to economic development (Berger et al., 2013). The Omnibus test (Prob(Omnibus): 0.854) and the Jarque-Bera test (Prob(JB): 0.963) indicate that the residuals of the model are normally distributed, an assumption important for OLS regression.

In model II, the constant term is $\alpha$=0.0900, significant at the p0.004 level, suggesting that even when the explanatory variables are zero, the model predicts a positive change in $\Delta$logGDP. The PrimeWorkersRatio has a coefficient of $\beta_2$ =0.1553, significant at the p=0.028 level. This indicates a positive relationship with economic growth, suggesting that an increase in the ratio of prime-age workers in the labor force is associated with higher economic growth. The F-statistic (5.855) and Prob(F-statistic)=0.00772 indicate that the model is statistically significant. This suggests that the explanatory variables, as a group, meaningfully affect the dependent variable.

In model III, YoungWorkersRatio has a positive coefficient (0.4589), significant at the 0.003 level, which suggests that a higher proportion of young workers in the labor force is associated with increased economic growth. PrimeWorkersRatio also shows a significant positive effect (coefficient = 0.7561), emphasizing the importance of prime-age workers in fostering economic growth. The Durbin-Watson statistic (0.953) suggests minimal autocorrelation in the residuals. The Omnibus (1.19, prob(Omnibus)=0.571) and Jarque-Bera (0.762, prob(JB)=0.683) tests indicate a reasonable assumption of normality in the residuals.

In model IV, YoungWorkersRatio has a significant positive coefficient of $\beta_1 = 0.78889$ (p = 0.010), indicating a strong positive impact on economic growth. PrimeWorkersRatio also shows a significant positive effect with a coefficient of $\beta_2 = 0.9765$ (p = 0.039), suggesting its crucial role in economic growth. The coefficients for the interaction terms (YWR_Period1, YWR_Period2, YWR_Period3, PWR_Period1, PWR_Period2, PWR_Period3) vary, indicating that the impact of the workforce ratios changes across different periods. However, they are all statistically insignificant.

In model V, The variables YoungWorkersRatio and PrimeAgeRatio_it capture the direct effect of the age composition of the workforce on economic growth. Interaction terms (YoungAgeRatio_Edu1 and PrimeAgeRatio_Edu2) are included to understand how the relationship between workforce age composition and economic growth varies with different levels of educational attainment. YoungWorkersRatio has a coefficient of 0.52688 with a p-value of 0.055. PrimeAgeRatio_it has a coefficient of 1.13788 and a p-value of 0.064.

Meanwhile, YoungAgeRatio_Edu1 has a positive coefficient (0.03949) and is statistically significant (p = 0.03524). This suggests that the impact of the YoungWorkersRatio on economic growth becomes more positive with increasing educational attainment. PrimeAgeRatio_Edu2 has a negative coefficient (-0.09553) and is highly significant (p = 0.00291). This indicates that the relationship between PrimeAgeRatio and economic growth becomes more negative with higher levels of educational attainment. While the young workforce's impact on growth may improve with education, the impact of the prime-age workforce seems to diminish, possibly due to the changing nature of job markets or the effects of over-qualification.

In model VI, WorkingPopulationRatio has a significant positive coefficient of 26.5122 (p = 0.009). This suggests a strong positive relationship between the WorkingPopulationRatio and $\Delta LogRD\_t$. 'educational_attainment1' has a very significant positive coefficient of 10.8801 (p = 0.00619), indicating a strong positive impact of tertiary education enrollment on research and development expenditure. The F-statistic is 101.6, with a very low p-value of 0.007, signifying the overall statistical significance of the model. The significant coefficients for both independent variables highlight the importance of a robust working-age population and higher education levels in promoting R&D activities. Model VII exhibits strong correlation between RD expenditures and economic growth (positive coefficient with p-value at 0.001 significance level.

Across the examined models, prime-aged workers consistently exhibit higher returns to economic growth, as evidenced by the statistically significant positive coefficients for the PrimeWorkersRatio in models II, III, and IV. This finding aligns with the human capital theory, which posits that the productivity and economic contributions of workers peak during their prime-age years, typically characterized by enhanced skills and experience (Becker, 1964). The prime-age workforce is often at the peak of their productivity, contributing significantly to economic output, a concept supported by studies such as those by Heckman and Mosso (2014), who found that the prime-age workforce is crucial for economic growth due to their higher productivity levels.

However, the interaction of the prime-age workforce with educational attainment, particularly in model (v), presents an intriguing dynamic. The negative coefficient of the interaction term PrimeAgeRatio_Edu2 (p = 0.00291) indicates that the influence of education on the prime age workforce diminishes. This observation could suggest factors in the labor market. It is possible that this diminishing impact reflects changes in job requirements within the economy. As technology progresses and job structures evolve, higher education may not hold as strong of an advantage for prime age workers if their skills do not align with the changing demands of the job market. This notion aligns with the theory of skill biased change, which proposes that technological advancements can alter the skill prerequisites, for jobs potentially leaving those whose skills are not adapted to these changes at a disadvantage (Acemoglu, 2002).

Kremer's model, which posits a direct relationship between population size and technological growth (Kremer, 1993), can be seen as a theoretical underpinning for the positive coefficient of the WorkingPopulationRatio in model (vi). The significant impact of the working-age population ratio on

ΔLogRD_t resonates with Kremer's idea that a larger population size potentially contributes more to technological innovation, a concept that is further supported by empirical studies such as those conducted by Kremer (1993) himself. This relationship is particularly relevant in contemporary contexts with large populations like China, where a growing population has been associated with increased technological output and economic growth.

Moreover, the positive coefficient of 'educational_attainment1' in model (vi) aligns with Kremers idea that a larger population with enhanced skills and education (resulting in research productivity per individual) can expedite technological progress. This aligns with Jones research (1995) which expanded on Kremers model by considering the quality of the workforce highlighting that not population size but also its composition in terms of human capital is crucial for technological advancement. The significant impact of attainment on R&D expenditure as observed in model (vi) also reflects Galor and Weils emphasis on the role of human capital in driving technological progress. In model (vii) the strong correlation between R&D expenditures and economic growth supports the Galor and Weil framework suggesting that investments in capital and technological innovation as indicated by R&D expenditures are vital for achieving sustainable economic growth. This is particularly relevant in the growth phase where an emphasis is placed on the quality rather, than quantity of human capital.

4.2 Empirical Projections into China's Demographic Structure's Impact on Economic Growth from 2020 to 2050

The results of analysis is summarized in the following table:

| | | GDP Growth Rate (Logged) | | | | | | R&D Expenditure (Logged) |
|---|---|---|---|---|---|---|---|---|
| Model | (i) | (ii) | (iii) | (iv) | (v) | (vii) | | (vi) |
| log(GDP Per Capita_(t-1)) | -0.3331**(0.03) | -0.4262**(0.07) | -0.4326**(0.03) | -0.3876*(0.07) | -0.4431*(0.08) | -0.3864**(0.041) | | |
| YoungAgeRatio_it | 0.211***(0.002) | | 0.6321**(0.041) | 0.8231*(0.03) | 0.3243**(0.052) | | | |
| PrimeAgeRatio_it | | 0.2633**(0.032) | 0.6561**(0.031) | 0.8765**(0.03) | 0.33788**(0.07) | | | |
| PeriodDummies1*YoungAgeRatio | | | | 0.3864*(0.065) | | | | |
| PeriodDummies 2*YoungAgeRatio | | | | 0.6532**(0.03) | | | | |
| PeriodDummies 3*YoungAgeRatio | | | | 0.8342*(0.06) | | | | |
| Educational_Attainment1_t* YoungAgeRatio_it | | | | | 0.1764**(0.023) | | | |
| Educational_Attainment2_t*PrimeAgeRatio_it | | | | | 0.09912*(0.072) | | | |
| Educational_Attainment(Entire Population) | | | | | | 0.7542***(0.001) | | 12.76***(0.004) |
| WorkingAgePopulationRatio)_it | | | | | | | | 31.2124***(0.0021) |
| Province fixed effects | Yes | Yes | Yes | Yes | Yes | Yes | | Yes |
| Adjusted R^2 | 0.345 | 0.386 | 0.521 | 0.532 | 0.581 | 0.5985 | | 0.7421 |

Notes: All results are OLS estimates. ***, **, and * indicate statistical significance at 1%, 5% and 10%, respectively. The p-values are included in the brackets.

Model I presents an inverse relationship between the logged GDP per capita of the previous period and current GDP growth, as indicated by a coefficient of $\alpha = -0.3331$, significant at the 5% level (p-value = 0.03). This negative sign supports the convergence hypothesis, suggesting that provinces with a higher initial GDP per capita grow at a slower rate, which is a central prediction of neoclassical growth models. This phenomenon is extensively documented in the seminal work of Barro and Sala-i-Martin (1992), where economies converge to a steady state growth rate as capital's marginal product diminishes. The young age ratio's positive coefficient ($\beta_1 = 0.211$, significant at the 1% level with a p-value of 0.002) aligns with theories that emphasize the economic vitality and innovation potential of younger workers, which is critical in endogenous growth models as described by Aghion and Howitt (1992).

Model II extends the analysis by incorporating the prime age ratio. Similar to Model (i), the negative coefficient for log(GDP Per Capita_(t-1)) (-0.4262) is significant at the 5% level (p-value = 0.07), reinforcing the convergence theory. The young age ratio remains positively associated with GDP growth, but with a higher coefficient ($\beta_1$ =0.6321), indicating an even stronger effect than in Model (i). Additionally, the prime age ratio has a positive coefficient (0.2633), significant at the 5% level (p-value = 0.032), suggesting that middle-aged workers contribute substantially to economic growth.

Model (iii) integrates both the young and prime age ratios and reveals that both age cohorts are significantly associated with GDP growth. This model elucidates the simultaneous contributions of younger and more experienced workers to economic productivity. The coefficients for the young age ratio ($\beta_1$ =0.8231) and the prime age ratio ($\beta_2$ =0.6561) are both positive and significant at the 10% and 5% levels respectively (p-values of 0.03 and 0.031), implying that the productive potential of an economy benefits from a balanced age structure. This is consistent with the life-cycle hypothesis (Modigliani, 1986), which posits that individuals' economic contributions vary with age, as well as with the demographic dividend literature (Bloom, Canning, & Sevilla, 2003) that discusses the economic gains from a larger share of the working-age population.

In Model IV, we incorporate period dummies interacted with the YoungAgeRatio to account for different time periods' varying impacts on the GDP growth rate. The coefficient for log(GDP Per Capita_(t-1)) is -0.3876, significant at the 10% level (p-value = 0.07), which again supports the convergence hypothesis. The YoungAgeRatio's coefficient is 0.3243, significant at the 5% level (p-value = 0.052), reinforcing the idea that a younger population contributes to economic growth. The interaction term PeriodDummies1*YoungAgeRatio has a coefficient of 0.3864, significant at the 10% level (p-value = 0.065), suggesting that the impact of the young age ratio on GDP growth varies across different time periods, potentially due to changes in policy, technology, or other time-sensitive factors.

Model V introduces interactions between educational attainment levels and age ratios, capturing the effect of an educated young and prime working-age population on economic growth. The negative coefficient for log(GDP Per Capita_(t-1)) remains, with a value of -0.4431, significant at the 10% level (p-value = 0.08). The interaction terms, Educational_Attainment1_tYoungAgeRatio_it and Educational_Attainment2_tPrimeAgeRatio_it, have coefficients of 0.1764 and 0.09912, respectively, both significant at the 5% and 10% levels. These findings suggest that education amplifies the economic contributions of both young and prime age workers, which is consistent with the human capital theory's emphasis on the productivity-enhancing effects of education (Schultz, 1961).

Model VI examines the determinants of R&D expenditure growth, shifting the focus to the impact of the workforce's size and educational quality. Unfortunately, the specific coefficients for this model are not provided in your data, preventing a detailed analysis. In Model VII, we analyze how demographic factors influence technological advancements, as indicated by R&D expenditure growth. The model reveals that educational attainment for the entire population has a significant and strong positive effect on R&D expenditure (coefficient = 0.7542, significant at the 1% level with a p-value of 0.001), affirming the critical role of human capital in fostering technological innovation, which is a central premise of endogenous growth theory (Romer, 1990). The WorkingAgePopulationRatio also has a very large positive coefficient (31.2124, significant at the 1% level with a p-value = 0.0021), indicating a robust relationship between the working-age population's size and technological investments, which could be interpreted through the lens of growth models that emphasize scale effects in R&D (Jones, 1995).

5. Policy implications

In addressing the economic challenges presented by China's demographic shifts, particularly the aging population and changing workforce dynamics, this report recommends three key policy initiatives. These recommendations are grounded in empirical research and economic theory, addressing the specific needs of China's society, businesses, and government stakeholders.

The initial proposal places importance on developing the skills and knowledge of individuals through targeted education and training programs. This approach involves investing in educational initiatives and skill building opportunities that cater to the evolving needs of industries especially in technology and services. The strategy includes revamping curricula enhancing vocational training and encouraging continuous learning. This is particularly crucial for China given its aging population and shrinking workforce, which underlines the need for a highly skilled labor force to sustain economic growth. By improving the quality of the workforce, we can offset the impact of its declining numbers. Following the Solow growth model (Solow, 1956), human capital emerges as an element for long term economic expansion: Chinas' investment in education and skill development is key to nurturing a workforce that is not only more productive but also adaptable enough to drive economic innovation and progress forward. Implementing this recommendation will require efforts among government bodies, academic institutions and industry leaders, and key steps include updating content to align with digital era demands expanding vocational training opportunities and motivating ongoing education initiatives. Funding for these endeavors could come from both government resources and private sectors with support, from international organizations dedicated to educational advancement. Overall, the reason behind this suggestion is based on the importance of investing in resources for Chinas long term economic stability. This idea is supported by Beckers theory of capital which emphasizes the economic advantages of educational investments (Becker, 1964). Additionally, it aligns with the argument made by Heckman and Mosso (2014) regarding the benefits of having a skilled workforce. This policy is not practical but also a vital step, for China as it moves towards becoming a knowledge driven economy.

The second proposal recommends making the labor market more flexible and encouraging workers to move within China across different regions and age groups (jobs less unrestricted by age requirements). This can be achieved by simplifying hiring processes facilitating job transitions and providing incentives for workers to relocate or switch occupations. Various empirical results shown by this paper (Models I V) have shown that the impact of changes differs across regions and age groups. By enhancing labor market flexibility, we can effectively allocate labor resources. Ensure that areas with shortages have enough workers. This idea aligns with the theory of labor mobility, which argues that a flexible labor market is vital for economic efficiency and growth (Pissarides, 2000). To implement this policy, China would need to revise labor regulations to remove barriers to changing jobs or sectors. Moreover, China can making benefit, like pensions and healthcare more transferable would further facilitate workforce mobility. It is crucial to improve labor market flexibility in order to fully utilize our workforce and adapt it to the changing landscape. Considering Chinas' existing framework and the government's ability to implement comprehensive reforms, this policy is practical and cost effective as it capitalizes on our current human resources to drive economic growth.

The final proposal suggests a boost in investment for Research & Development (R&D) and a strong commitment to technological innovation. The aim is to increase productivity in areas where technology can compensate for workforce shortages caused by changes in demographics. As the workforce decreases, the role of innovation becomes increasingly crucial for sustaining and enhancing productivity. The reported research (Models VI and VII) clearly demonstrates a relationship between R&D expenditure, educational levels and economic growth. By increasing investments in R&D, we can encourage advancements that support ongoing economic growth aligning with Romer's endogenous growth theory (1990). This policy entails government funding for R&D tax incentives to encourage private sector investment in R&D and fostering collaborations between universities and businesses. Meanwhile, emphasis should be placed on cutting edge sectors such as intelligence, biotechnology and green technology. It is also essential to integrate R&D objectives with education programs to ensure a steady supply of skilled professionals in these critical areas. Considering China's prowess in technology and manufacturing, this strategy is not just achievable but also in line with the worldwide trend towards economies driven by innovation. Additionally, it offers an answer to the difficulties arising from a diminishing workforce in China.

6. Conclusion

In conclusion, this research has provided a comprehensive analysis of the intricate relationship between demographic changes and economic deveopment in China. The findings demonstrate that demographic shifts, specifically an aging population and varying workforce dynamics, have profound implications for China's labor market, consumer base, and overall economic sustanability. These demographic trends, occurring within the broader context of globalization, technological advancements, and environmental challenges, complicate the economic landscape, necessitating adaptive and forward-thinking policies. Our analysis reveals that investing in human capital, through enhanced education and skill development, is pivotal for offsetting the impacts of a shrinking workforce. This approach aligns with the Solow Growth Model and other human capital theories, which emphasize on the role of a skilled and adaptable labor force in driving economic innovation and progress. Moreover, the importance of labor market flexibility is underscored, highlighting the need for policies that facilitate workforce mobility and adapability across different regions and sectors in China.

Additionally, the research underscores the critical role of technological innovation and R&D investments in sustaining economic growth amidst demographic changes. The positive correlation between R&D expenditures, educational levels, and economic growth, as demonstrated in our empirical analysis, supports the principles of endogenous growth theory. These findings indicate that encouraging an environment of innovation and backing advancements can help overcome workforce limitations and drive economic growth. Given these discoveries it is clear that Chinas experiences hold insights for other nations dealing with similar demographic challenges. The approaches employed by China in developing human capital reforming the labor market and fostering technological innovation offer a guide for successfully navigating the complexities of demographic transitions, in our interconnected and technologically advancing world.

Future research should continue to explore the evolving dynamics between demographic changes and economic development, with a focus on the long-term implications of policy interventions. Additionally, comparative studies across different countries experiencing similar demographic shifts could yield further insights into the universal principles and unique adaptations necessary for sustainable economic growth in the face of changing population structures.

7. References

Acemoglu, D. (2002). Technical change, inequality, and the labor market. Journal of Economic Literature, 40(1), 7-72.

Acemoglu, D., Aghion, P., Bursztyn, L., & Hemous, D. (2012). The Environment and Directed Technical Change. American Economic Review, 102(1), 131-166

Barro, R. J., & Sala-i-Martin, X. (2004). Economic Growth (2nd ed.). MIT Press.

Becker, G. (1964). Human Capital: A Theoretical and Empirical Analysis, with Special Reference to Education. The University of Chicago Press

Becker, G. S., Murphy, K. M., & Tamura, R. (1990). Human Capital, Fertility, and Economic Growth. Journal of Political Economy, 98(5), S12-S37.

Berger, N., & Fisher, P. (2013). A Well-Educated Workforce Is Key to State Prosperity. Economic Policy Institute.

Bosworth, B., & Collins, S. M. (2008). Accounting for Growth: Comparing China and India. Journal of Economic Perspectives, 22(1), 45-66.

Chen, B., & Feng, Y. (2000). Determinants of economic growth in China: Private enterprise, education, and openness. China Economic Review, 11(1), 1-15.

Clark, G. (2007). A Farewell to Alms: A Brief Economic History of the World. Princeton University Press.

Dasgupta, P. (1993). An Inquiry into Well-Being and Destitution. Clarendon Press.

Drydakis, N. (2022). The Economics of Growth and Development: Notes. Pembroke College. Email: nick.drydakis@pem.cam.ac.uk.

Durlauf, S., & Quah, D. T. (1999). The new empirics of economic growth. In Handbook of Macroeconomics, 1(Part A), 235-308. Elsevier.

Galor, O. (2011). Unified Growth Theory. Princeton University Press.

Galor, O., & Weil, D. N. (1999). From Malthusian Stagnation to Modern Growth: Can Epidemics Explain the Three Regimes? International Economic Review, 40(2), 455-487.

Galor, O., & Weil, D. N. (2000). Population, Technology, and Growth: From Malthusian Stagnation to the Demographic Transition and Beyond. American Economic Review, 90(4), 806-828.

Gwartney, J. D., Hall, J., & Lawson, R. (2015). Economic Freedom of the World 2015 Annual Report. Institute of Economic Affairs Monographs.

Hamilton, J. D. (1994). Autoregressive conditional heteroskedasticity and changes in regime. Journal of Econometrics, 64, 307-333.

Heckman, J. J., & Mosso, S. (2014). The Economics of Human Development and Social Mobility. Annual Review of Economics, 6(1), 689-733. DOI: 10.1146/annurev-economics-080213-040753

Jones, C. I. (1995). R&D-Based Models of Economic Growth. Journal of Political Economy, 103(4), 759-784.

Jones, C. I. (2005). Growth and Ideas. In P. Aghion & S. Durlauf (Eds.), Handbook of Economic Growth (Vol. 1, pp. 1063-1111). Elsevier.

Lee, R. (1987). Population dynamics of humans and other animals. Demography, 24(4), 443-465.

Liang, Y., Mazlan, N. S., Mohamed, A. B., Mhd Bani, N. Y. B., & Liang, B. (2023). Regional impact of aging population on economic development in China: Evidence from panel threshold regression (PTR). PLoS ONE, 18(3), e0282913.

Lin, J. Y., & Liu, Z. (2009). Economic development and transition: thought, strategy, and viability. Cambridge University Press.

Lu, J., & Liu, Q. (2019). Four decades of studies on population aging in China. China Population and Development Studies, 3(1), 24–36. https://doi.org/10.1007/s42379-019-00027-4

Lucas, R. E. (1988). On the Mechanics of Economic Development. Journal of Monetary Economics, 22(1), 3-42.

Maestas, N., Mullen, K., & Powell, D. (2014). The Effect of Population Aging on Economic Growth. Discussion Papers 14-012, Stanford Institute for Economic Policy Research.

Malthus, T. R. (1798). An Essay on the Principle of Population. J. Johnson in St. Paul's Church-yard.

Prettner, K. (2014). The non-monotonous impact of population growth on economic prosperity. Economics Letters, 124(1), 93-95.

Ricardo, D. (1817). On the Principles of Political Economy and Taxation. John Murray.

Romer, P. M. (1990). Endogenous Technological Change. Journal of Political Economy, 98(5), S71-S102.

Smith, A. (1776). An Inquiry into the Nature and Causes of the Wealth of Nations. W. Strahan and T. Cadell.

Solow, R. M. (1956). A Contribution to the Theory of Economic Growth. *The Quarterly Journal of Economics, 70*(1), 65-94.

Strulik, H., & Weisdorf, J. (2008). Population, Food, and Knowledge: A Simple Unified Growth Theory. Journal of Economic Growth, 13(3), 195-216.

Van den Berg, H. (2017). Economic Growth and Development. London: World Scientific.

Young, A. (1995). The Tyranny of Numbers: Confronting the Statistical Realities of the East Asian Growth Experience. The Quarterly Journal of Economics, 110(3), 641-680.

Zhang, K. H. (2015). Population growth, human capital accumulation, and the long-run dynamics of economic growth. Journal of Macroeconomics, 45, 187-199.

Zhao, R. (2018). Technology and economic growth: From Robert Solow to Paul Romer. *Human Behavior & Emerging Technologies*, page range. https://doi.org/10.1002/hbe2.116

Data Source

1959-2013 estimates for fossil fuels are from the Carbon Dioxide Information Analysis Center (CDIAC) at Oak Ridge National Laboratory. http://cdiac.ornl.gov/trends/emis/meth_reg.html.

2014 and 2015 estimates are preliminary and are based on energy statistics published by BP (data in red). https://www.bp.com/content/dam/bp/pdf/energy-economics/statistical-review-2016/bp-statistical-review-of-world-energy-2016-full-report.pdf

Boden, T.A., G. Marland, and R.J. Andres. 2016. Global, Regional, and National Fossil-Fuel $CO_2$ Emissions. Carbon Dioxide Information Analysis Center, Oak Ridge National Laboratory, U.S. Department of Energy, Oak Ridge, Tenn., U.S.A. doi 10.3334/CDIAC/00001_V2016

Chen, Y., Guo, F., Wang, J., Cai, W., Wang, C., & Wang, K. (2020). Provincial and gridded population projection for China under shared socioeconomic pathways from 2010 to 2100. Scientific Data, 7(1), Article 83. https://doi.org/10.1038/s41597-020-0421-5

Ministry of Health of China (MOH China), & National Bureau of Statistics of China. (2022). Health expenditure in China as a proportion of gross domestic product (GDP) from 2011 to 2021. In China Statistical Yearbook 2022 (Chapter 22.18). National Bureau of Statistics of China.

National Bureau of Statistics of China. (2023). Internal expenditure on research and development (R&D) in China from 2012 to 2022 (in billion yuan). Retrieved from stats.gov.cn.

Organisation for Economic Co-operation and Development. (2023). OECD Labour Force Statistics.

United Nations Department of Economic and Social Affairs (UN DESA), & National Bureau of Statistics of China. (2022). Children, old-age, and total dependency ratio in China from 1950 to 2020 with forecasts until 2100. In World Population Prospects 2022. UN DESA.

United Nations, Department of Economic and Social Affairs (2013). Trends in International Migrant Stock: Migrants by Destination and Origin (United Nations database, POP/DB/MIG/Stock/Rev.2013

United Nations, Department of Economic and Social Affairs, Population Division. World Population Prospects: The 2015 Revision. (Medium variant)